\documentclass[aps,superscriptaddress,nobibnotes,prb,twocolumn]{revtex4}
\usepackage{graphicx}
\usepackage{dcolumn}
\usepackage{bm}
\usepackage{amsmath}
\usepackage{textcomp}
\usepackage{color}
\usepackage[squaren]{SIunits}
\usepackage{natbib}
\usepackage{tabularx}

\begin{document}

\newcommand{\ie}{{\it i.e.}}
\newcommand{\eg}{{\it e.g.}}
\newcommand{\etal}{{\it et al.}}


\title{Evolution of transport properties in FeSe thin flakes with thickness approaching the two-dimensional limit}

\author{C. S. Zhu$^{1}$, B. Lei$^{1}$, Z. L. Sun$^{1}$, J. H. Cui$^{1}$, M. Z. Shi$^{1}$, W. Z. Zhuo$^{1}$, X. G. Luo}
\altaffiliation{E-mail: xgluo@ustc.edu.cn}
\affiliation{Key Laboratory of Strongly Coupled Quantum Matter Physics, Chinese Academy of Sciences, Hefei National Laboratory for Physical Sciences at Microscale, Department of Physics, University of Science and Technology of China, Hefei 230026, China}

\author{X. H. Chen}
\altaffiliation{E-mail: chenxh@ustc.edu.cn}
\affiliation{Key Laboratory of Strongly Coupled Quantum Matter Physics, Chinese Academy of Sciences, Hefei National Laboratory for Physical Sciences at Microscale, Department of Physics, University of Science and Technology of China, Hefei 230026, China}
\affiliation{CAS Center for Excellence in Superconducting Electronics (CENSE), Shanghai 200050, China}
\affiliation{CAS Center for Excellence in Quantum Information and Quantum Physics, Hefei, Anhui 230026, China}


\begin{abstract}
Electronic properties of FeSe can be tuned by various routes. Here, we present a comprehensive study on the evolution of the superconductivity and nematicity in FeSe with thickness from bulk single crystal down to bilayer ($\sim$ 1.1 nm) through exfoliation. With decreasing flake thickness, both the structural transition temperature $T_{\rm s}$ and the superconducting transition temperature $T_{\rm c}^{\rm zero}$ are greatly suppressed. The magnetic field ($B$) dependence of Hall resistance $R_{xy}$ at 15 K changes from $B$-nonlinear to $B$-linear behavior up to 9 T, as the thickness ($d$) is reduced to 13 nm. $T_{\rm c}$ is linearly dependent on the inverse of flake thickness (1/$d$) when $d\le$ 13 nm, and a clear drop of $T_{\rm c}$ appears with thickness smaller than 27 nm. The $I$-$V$ characteristic curves in ultrathin flakes reveal the signature of Berezinskii-Kosterlitz-Thouless (BKT) transition, indicating the presence of two-dimensional superconductivity. Anisotropic magnetoresistance measurements further support 2D superconductivity in few-layer FeSe. Increase of disorder scattering, anisotropic strains and dimensionality effect with reducing the thickness of FeSe flakes, might be taken into account for understanding these behaviors. Our study provides systematic insights into the evolution of the superconducting properties, structural transition and Hall resistance of a superconductor FeSe with flakes thickness and  provides an effective way to find two-dimensional superconductivity as well as other 2D novel phenomena.\\

\end{abstract}

\maketitle

\section{Introduction}
Iron-based superconductors have triggered widespread interest since they were discovered to possess high transition temperature \cite{Z1,X2} and high upper critical fields \cite{F3,H4,H5}. Because of its simplest structure and high tunability among the iron-based superconductors, FeSe is a good platform for studying the pairing mechanism of superconductivity and the correlation between different electronic orders\cite{F6}. Although bulk FeSe only exhibits superconductivity about 8 K \cite{K7}, its $T_{\rm c}$ can be significantly enhanced by electron doping \cite{B8}, external pressure \cite{J9}, and organic ion intercalation \cite{M10}. In particular, the monolayer FeSe film on a SrTiO$_3$ substrate has generated wide research interest because of its unexpected high-$T_{\rm c}$ superconductivity close to the boiling temperature of liquid nitrogen \cite{Q11,Y12}. Such a sharply contrasting $T_{\rm c}$ between bulk and monolayer specimen suggests that the interface between FeSe and SrTiO$_3$ should be the key for the dramatic enhancement of superconductivity. Therefore, it is significant to find out how the $T_{\rm c}$ of the FeSe changes without the interface between FeSe and substrates as it is thinned towards the monolayer limit, since there could exist weak interactions between flakes and substrates. It has been reported that superconductivity is usually suppressed as the thickness of the FeSe flakes is reduced, and resistance cannot even reach zero as cooled down to 2 K for flakes with thickness of 9 - 10 nm \cite{B8,L13}. However, the FeSe thin film grown on bilayer graphene shows superconductivity (characterized by $in~situ$ scanning tunneling microscopy) even when the thickness is reduced to bilayer\cite{C14}. Therefore, we can expect to observe superconductivity in flakes thinner than 9 nm by adopting more effective way to protect the flake devices. FeSe film grown on bilayer graphene is thought to be strain-free. But at a usual situation in approaching 2D limit, strain from substrate has to be considered as an important factor to affect electronic states. It is well known that strain is a powerful tool to tune electronic properties in FeSe \cite{F15,M16}. The strain is mainly induced by mechanically external force or the mismatch in the lattice parameters between the film and the substrate. For instance, FeSe film grown on CaF$_2$ is compressive-strained, but that grown on SrTiO$_3$ is tensile-strained. The tensile strain reduces the superconducting transition temperature and enhances the structural transition temperature. On the contrary, the compressive strain enhances superconductivity but it suppresses the nematic phase \cite{G17}. This seems to suggest that the nematic phase is competing with superconductivity. When FeSe flakes is exfoliated onto a substrate (other than bilayer graphene) and the thickness of flakes approaches 2D limit, strains might emerge due to the difference of thermal expansion coefficients between ultrathin flakes and substrates \cite{R18}, so that in addition to reduction of dimensionality, possibly existing strains could also have effect on structural and superconducting transition.

A challenging issue is to achieve ultrathin FeSe flake. As one knows, although bulk FeSe can be cleaved along the van der Waals gap, it is difficult to get the few-layer sample via the traditionally mechanical exfoliation. Recently, an Al$_2$O$_3$-assisted exfoliation technique was successfully used for fabricating atomically thin MnBi$_2$Te$_4$ and Fe$_3$GeTe$_2$ flakes \cite{Y18,Y19}. Therefore, we try to use this method to get few-layer FeSe flakes.

In this paper, we use traditionally mechanical and Al$_2$O$_3$-assisted exfoliation techniques to obtain the thin-flake FeSe with different thicknesses. Upon cooling, bulk FeSe undergoes tetragonal-orthorhombic structural transition at $T_{\rm s}$ $\sim$ 90 K \cite{T20}, accompanied by a superconducting transition at $T_{\rm c}^{\rm zero}$ $\sim$ 8 K \cite{A21}. As the thickness is reduced from bulk towards few-layers, we observe the reduction of both $T_{\rm s}$ and $T_{\rm c}$. $T_{\rm s}$ decreases from $\sim$ 90 K in bulk to $\sim$ 57 K in the pentalayer flake, and structural transition cannot be observed in resistance in the tetralayer flake. $T_{\rm c}^{\rm zero}$ changes from 8 K in bulk to 2.4 K in the tetralayer flake. The trilayer FeSe shows an upturn in resistance below $\sim$ 70 K, and subsequently a sharp decrease near 3.6 K with further cooling, indicating a superconducting transition with $T_{\rm c}^{\rm onset}$ $\sim$ 3.6 K. The bilayer FeSe just displays an insulating behavior without any transition in resistance down to 2 K. With decreasing the flake thickness, the magnetic field ($B$) dependence of Hall resistance $R_{xy}$ at 15 K changes from $B$-nonlinear to $B$-linear dependence up to 9 T. Finally, the $R_{xy}$ at 15 K clearly shows perfect $B$-linear dependence up to 9 T in flakes with $d\le$ 13 nm. Furthermore, the $I$-$V$ characteristic curves in hexalayer flake reveal the signature of BKT transition, which is an evidence for 2D superconductivity in few layer FeSe \cite{W22,B23.1}. Anisotropic magnetoresistance measurements further support 2D superconductivity in few-layer FeSe. Increase of disorder scattering and anisotropic strains in the nematic state, as well as reduction of dimensionality, might be taken into account for understanding these behaviors.

\section{Experimental Section}
Single crystals of pristine tetragonal FeSe were grown by a KCl-AlCl$_3$ flux method \cite{M24}. To obtain FeSe thin flakes with the thickness above 5 nm (including 5 nm), we directly use traditionally mechanical  exfoliation technique \cite{B8}. Single crystal pieces were mechanically exfoliated by using the scotch tape and then transferred onto the Si/SiO$_2$ substrate. An Al$_2$O$_3$-assisted exfoliation technique was used to get FeSe flakes thinner than 5 nm. For the Al$_2$O$_3$-assisted exfoliation technique, the Al$_2$O$_3$ film was firstly deposited by thermally evaporating onto a freshly prepared surface of the bulk crystals and then a thermal release tape was used to pick up the Al$_2$O$_3$ film, along with pieces of FeSe microcrystals separated from the bulk. The Al$_2$O$_3$/FeSe stack was subsequently released onto a piece of transparent polydimethylsiloxane (PDMS) film. Finally, the PDMS/FeSe/Al$_2$O$_3$ assembly was stamped onto a substrate and quickly peeled away the PDMS film, leaving the Al$_2$O$_3$ film covered with freshly cleaved FeSe flakes on the Si/SiO$_2$ substrate, and then Au(100 nm)/Cr(5 nm) electrodes were patterned onto an isolated thin flake for transport measurements. FeSe thin flake is very sensitive to the air, so all the device-fabrication processes were performed in an argon-filled glove box, where water and oxygen content are maintained below 0.1 ppm to avoid sample degradation. Once the device fabrication is completed, we sealed the device in a chip carrier with vacuum grease and a cover glass inside the glove box, and then transferred the whole package into a commercial Quantum Design physical property measurement system (PPMS) \cite{Y25} for carrying out the transport properties.

\section{Results and discussion}

\begin{figure}[bt]
\includegraphics[width = 0.48\textwidth]{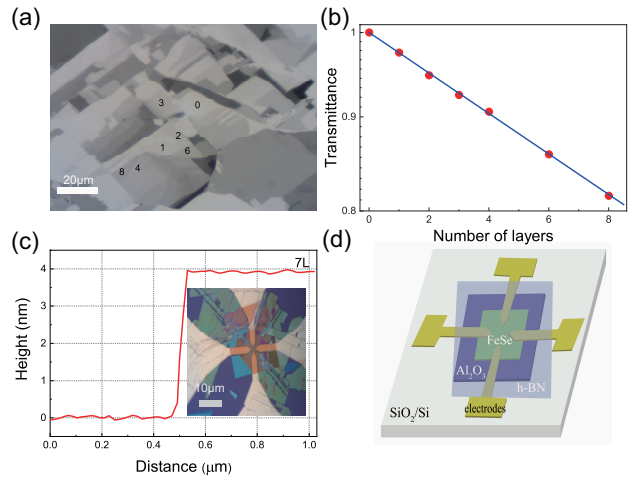}
\caption{(Color online) (a) Optical image of few-layer flakes of FeSe cleaved onto thermally evaporated Al$_2$O$_3$ thin film (thickness $\sim$ 80 nm). FeSe/Al$_2$O$_3$ stack is supported on a PDMS substrate. Image was taken in transmission mode. Number of layers is labeled on selected flakes. Scale bar, 20 $\mu$m. (b) Transmittance as a function of the number of layers. The transmittance (red dots) follows the Beer-Lambert law (blue line). (c) The thickness of a 7 layer FeSe is about 4 nm. The inset shows an optical image of a 7 layer FeSe device capped with a thin layer ($\sim$ 15 nm) of $h$-BN. Scale bar, 10 $\mu$m. (d) A schematic illustration of the FeSe thin flake device.}
\end{figure}

\begin{figure*}[bt]
\includegraphics[width = 0.85\textwidth]{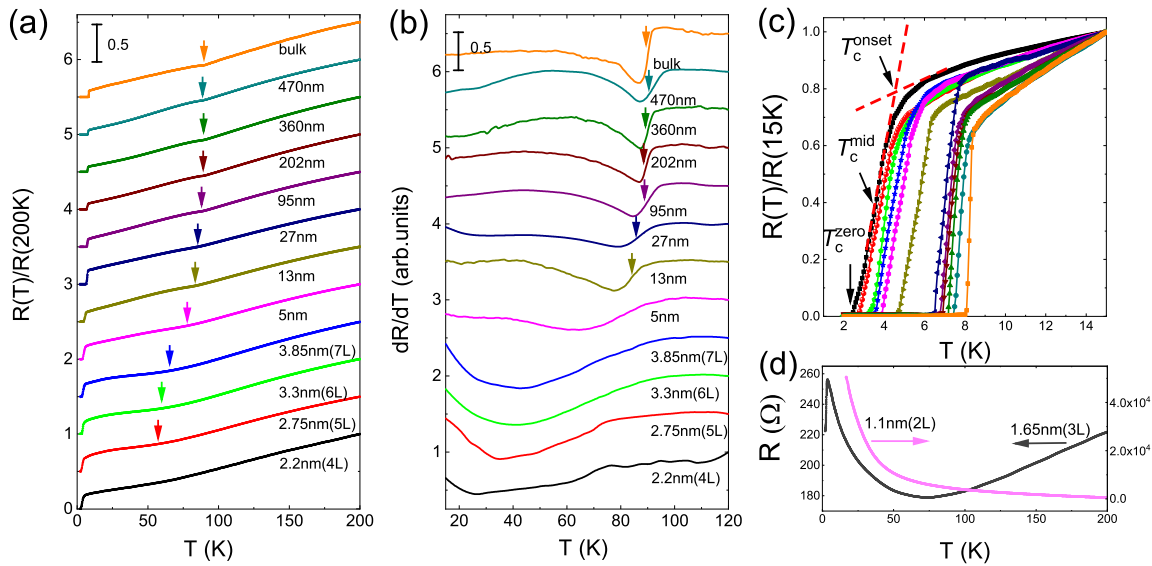}
\caption{(Color online)  (a) Temperature dependence of the normalized resistance $R$($T$)/$R$(200 K) for a bulk crystal and thin flakes with different thickness. The thinnest thickness is about 2.2 nm (4L). The arrows indicate the kink in the resistance curve, which corresponds to the structural transition ($T_{\rm s}$). (b) Temperature dependence of the normalized derivative of resistance, d$R$/d$T$, for the bulk crystal and thin flakes in (a). The arrows indicate the position of the structural transition, which is determined by the midpoint of the step-like anomaly of d$R$/d$T$. Note that the curves in (a) and (b) are shifted by 0.5 for clarity. (c) The details of normalized resistance $R$($T$)/$R$(15 K) with temperature range from 2 to 15 K for watching the superconducting transition. The $T_{\rm c}^{\rm onset}$ is determined from the cross point of two red dash lines. $T_{\rm c}^{\rm mid}$ is defined as the temperature at which resistance reaches half the value of normal state. $T_{\rm c}^{\rm zero}$ is defined as the temperature at which resistance reaches zero. The same color is used for the same thickness in (a), (b) and (c). (d) Temperature dependence of the resistance for bilayer and trilayer flakes.}
\end{figure*}

Using traditionally mechanical and Al$_2$O$_3$-assisted exfoliation techniques, we can obtain thin flakes of FeSe with different thicknesses. Figure 1a displays the optical image of few-layer FeSe flakes on the Al$_2$O$_3$ film attached to PDMS. The transmittance of various numbers of layers follows the Beer-Lambert law (Fig. 1b), which enables us to precisely determine the layer number \cite{Y18,Y19}. We use AFM to check the thickness of a 7-layer FeSe flake (the number of layers is decided by the transmittance). As shown in Fig. 1c, the thickness of 7-layer FeSe flake is about 4 nm, which is consistent with the thickness of 7 unit cell of FeSe, proving that the transmittance can totally determine the layer number. After depositing Cr/Au electrodes, the devices were capped with a thin-layer ($\sim$ 15 nm) h-BN (Fig. 1c inset) for protecting the FeSe flake. Fig. 1d shows a schematic illustration of the completed FeSe thin flake device.

\begin{figure}[bt]
\includegraphics[width = 0.48\textwidth]{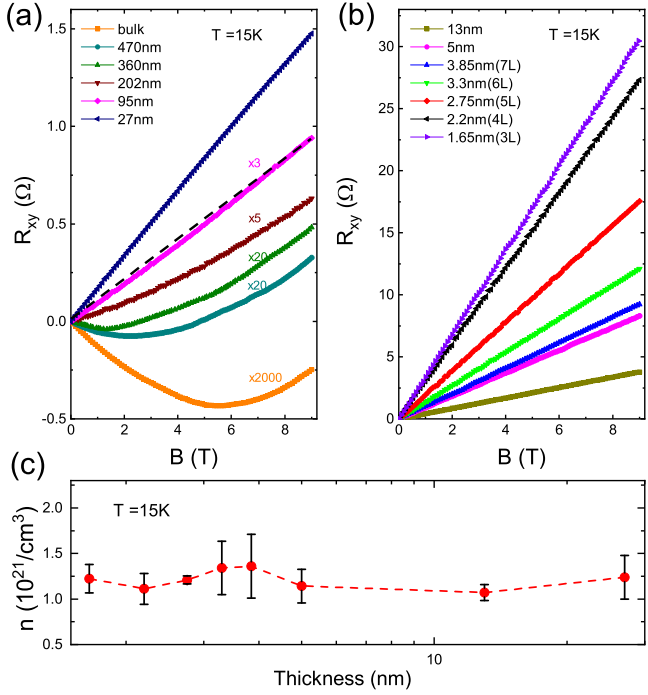}
\caption{(Color online) (a) and (b) The evolution of the magnetic-filed dependent Hall resistance $R_{xy}$ with different thickness at $T$ = 15 K. (c) Thickness dependence of carrier density $n$ at 15 K. The error bars are defined as the standard deviations of measurements.}
\end{figure}

\begin{figure}[bt]
\includegraphics[width = 0.48\textwidth]{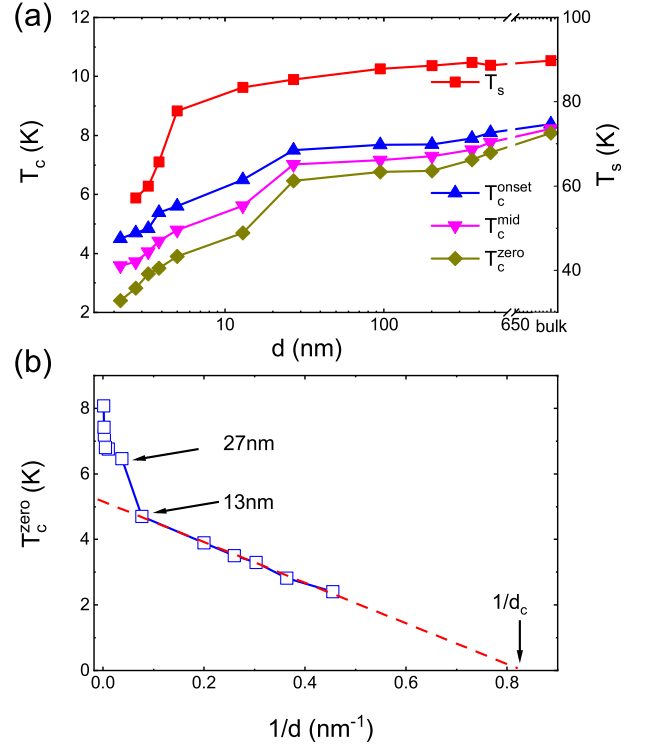}
\caption{(Color online) (a) Structural transition temperature and superconducting transition temperature as a function of thin flakes thickness $d$. (b) The $T_{\rm c}^{\rm zero}$ as a function of the inverse of the flakes thickness 1/$d$.}
\end{figure}

Figure 2a shows the typical temperature-dependence of the normalized resistance ($R$($T$)/$R$(200 K)) for a bulk crystal and various FeSe thin flakes with different thicknesses. The bulk FeSe shows a superconducting transition at $T_{\rm c}^{\rm zero}$ $\sim$ 8 K (where resistance reaches zero). With the thickness reduced from bulk towards 4 layers, $T_{\rm c}^{\rm zero}$ gradually changes from 8 to 2.4 K, as shown in Fig. 2c. $R$($T$) of the trilayer FeSe shows an upturn in resistance below $\sim$ 70 K, and subsequently a sharp decrease near 3.6 K with further cooling, indicating a superconducting transition with $T_{\rm c}^{\rm onset}$ $\sim$ 3.6 K. $R$($T$) of the bilayer FeSe displays an insulating behavior without any transition with temperature cooled down to 2 K (see Fig. 2d). A kink can be observed in the resistance curve, as displayed by the arrows in Fig. 2a, which is resulted from the structural transition from tetragonal to orthorhombic phase \cite{S26} (the corresponding temperature is denoted as $T_{\rm s}$) and indicates formation of a nematic phase upon cooling. $T_{\rm s}$ decreases with reducing thickness and the structural transition disappears for 4L flake. This can also be seen in the temperature derivative d$R$/d$T$ (as shown Fig. 2b) that the feature related to the structural transition becomes more and more obscure as thickness is reduced and eventually disappears as the flake is thinned to 4 layers. For each thickness of flakes, we repeated to fabricate more than three devices for measurements and obtained almost the same $T_{\rm c}^{\rm zero}$, $T_{\rm c}^{\rm onset}$ and $T_{\rm s}$.

\begin{figure*}[bt]
\includegraphics[width = 0.8\textwidth]{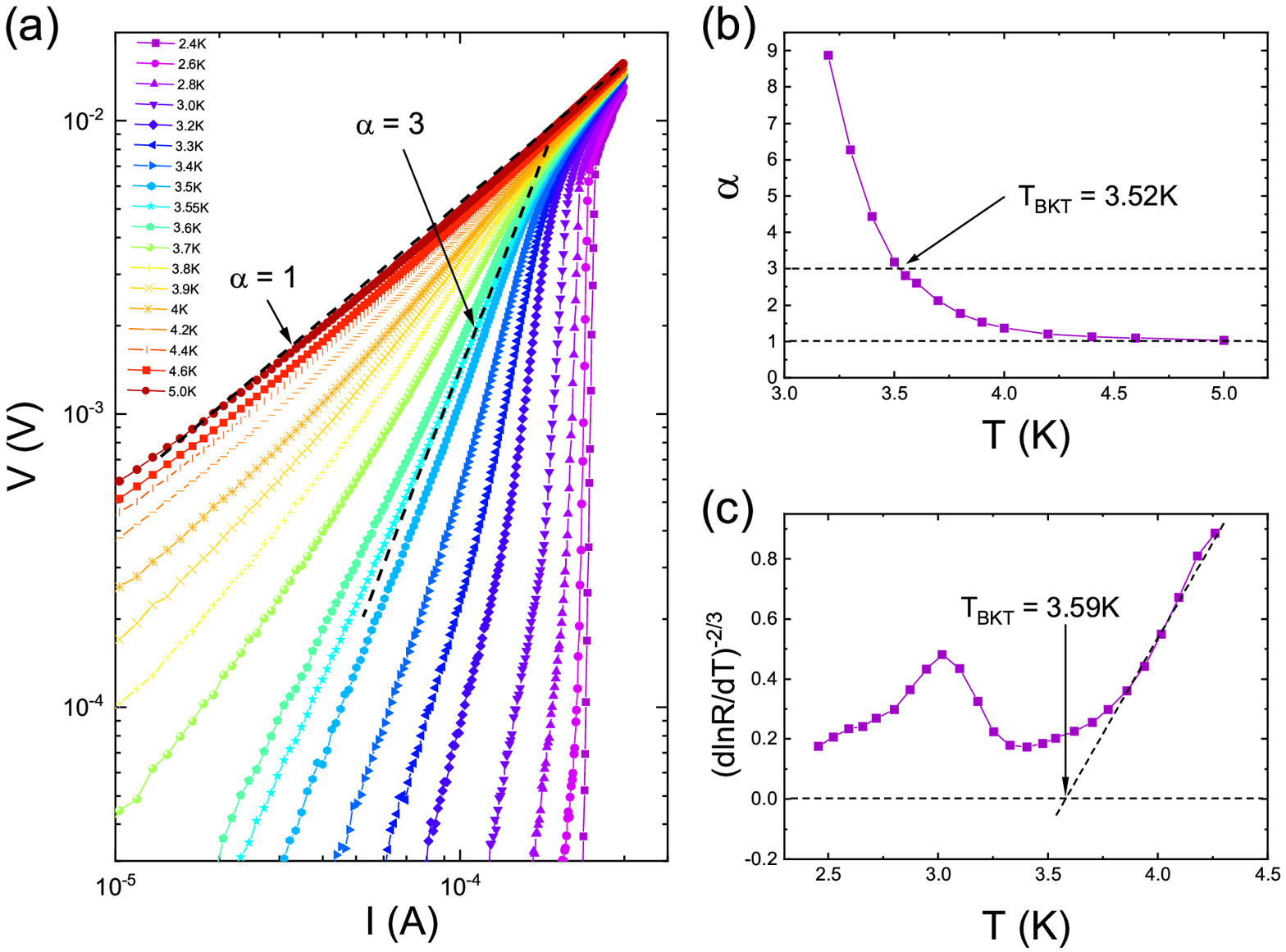}
\caption{(Color online) (a) The $V-I$ relationship at different temperatures in superconducting regime for hexalayer FeSe flake. The curves are plotted on logarithmic scale. The two black dashed lines refer to $V$ $\sim$ $I$ and $V$ $\sim$ $I^3$, respectively. (b) The variation of exponent $\alpha$ as a function of temperature, extracting from the power-law fittings in (a), showing that $T_{\rm BKT}$ = 3.52 K. (c) The $R({\rm T})$ curve plotted with the [dln($R$)/d$T$]$^{-2/3}$ scale. The dashed line shows the fitting to the Halperin-Nelson formula $R({\rm T})$ = $R_{\rm 0}$exp[$-b/(T - T_{\rm BKT})^{1/2}$] with $T_{\rm BKT}$ = 3.59 K.}
\end{figure*}

To further reveal the evolution of the electronic transport properties with decreasing the flakes thickness, we studied the Hall resistance $R_{xy}$ at 15 K for various FeSe flakes with different $d$. The $B$ dependence of $R_{xy}$ systematically changes with thickness. As shown in Fig. 3a and 3b, bulk FeSe shows a strong $B$-nonlinear dependence of Hall resistance at low temperature, which is attributed to a Dirac-type minor electron band with ultrahigh mobility \cite{H4,K7}. It is well known that FeSe is a multi-band system with several hole and electron band crossing Fermi level. It is reported that $R_{xy}$ is linear to magnetic field above $T_{\rm s}$ in bulk FeSe due to the compensation effect of holes and electrons \cite{K7,M24}, while becomes $B$-nonlinear below $T_{\rm s}$ because of the band reconstruction for the nematic transition, which leads to the change of the relative population of holes and electrons, and simultaneously arouses Dirac-like electrons with high mobility \cite{H4, K7, M24}. With decreasing flake thickness, the negative $R_{xy}$ is significantly suppressed and totally positive $R_{xy}$ is observed as thickness of the FeSe flake is reduced to 202 nm. With further thinning the FeSe flake to 13 nm, perfect positive $B$-linear dependence of $R_{xy}$ up to 9 T at 15 K can be observed. This systematically evolution of $R_{xy}$ at 15 K with the reduction of thickness seems to be consistent with rapid decrease of $T_{\rm s}$ as thickness is less than 27 nm. We directly extracted carrier density by using the relation $R_{\rm H}$ = ($R_{xy}$/$B$)$d$ (where $R_{xy}$ is the Hall resistance and $d$ is the sample thickness) and carrier density $n$ = 1/$e$$R_{\rm H}$. As shown in Fig. 3c, carrier density $n$ is almost a constant within the error bars for the thickness smaller than 27 nm.

\begin{figure}[bt]
\includegraphics[width = 0.48\textwidth]{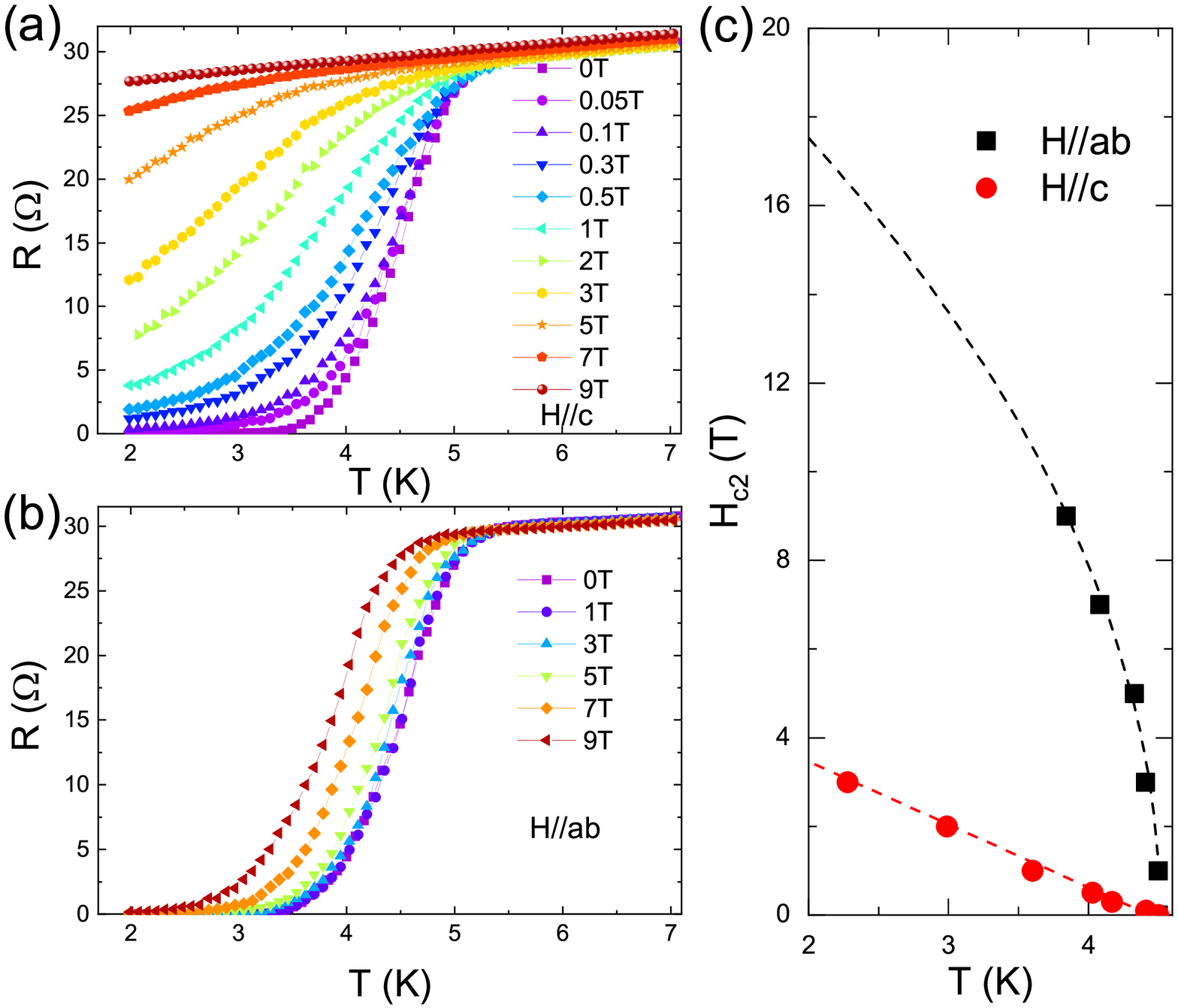}
\caption{(Color online) (a) and (b): Temperature-dependent resistance under a constant out-of-plane and in-plane magnetic field. (c) Temperature-dependent $H_{\rm c2}$ with magnetic field applied along out-of-plane (red) and in-plane (black) directions. Dashed curves are theoretical curves obtained from the 2D Ginzburg-Landau equations.}
\end{figure}

In order to clearly see the evolution of the structural transition as well as superconducting transition with the variation of thin flakes thickness, $T_{\rm s}$ and $T_{\rm c}$ are plotted as a function of thickness, as shown in Fig. 4a. When the thickness is greater than 27 nm, $T_{\rm s}$ and $T_{\rm c}$ are slightly suppressed. When the thickness is smaller than 27 nm, both $T_{\rm s}$ and $T_{\rm c}$ are significantly reduced. As shown in Fig. 4b, the $T_{\rm c}^{\rm zero}$ for the thin flakes with thickness $d\le$ 13 nm can be described by $T_{\rm c}$($d$)= $T_{\rm c0}$(1-$d_{\rm c}$/$d$) (where $T_{\rm c0}$ is superconducting transition temperature for bulk FeSe (infinite layers), $d$ is thickness, and $d_{\rm c}$ is the threshold for the emergence of superconductivity), which is successful for describing superconducting transition in ultrathin films, such as Pb, YBa$_2$Cu$_3$O$_y$ and FeSe \cite{M28,W29,C14}. Previous theoretical studies have shown that the relationship between $T_{\rm c}$ and the flakes thickness has successfully been interpreted by adding a surface-energy term in the Ginzburg-Landau free-energy of a superconductor \cite{J30}. For our flakes with thickness $d\le$ 13 nm, $T_{\rm c0}$ is extrapolated to be 5.14 K and a critical thickness $d_{\rm c}$ for losing superconductivity is 1.23 nm. This is consistent with the threshold for the onset of superconductivity that we experimentally observed in Fig. 2d (between 2 and 3 L).

Next we try to understand the results that both the $T_{\rm c}$ and $T_{\rm s}$ are suppressed with decreasing the flakes thickness. One possible origin for the simultaneous decrease of $T_{\rm s}$ and $T_{\rm c}$ with decreasing the thickness of the FeSe flakes could be the increase of disorder scattering with reducing thickness. For thin flakes, the size of twinned domains in the nematic state might decrease with reducing the thickness of the flakes, and more disorder scattering could be induced by the boundaries of domains, which effectively leads to decrease of $T_{\rm s}$ and $T_{\rm c}$ \cite{L13,A27}. In addition, the increase of disorder scattering arising from the boundaries of domains would dramatically reduce the mobility of the high-mobility Dirac-like electrons in the nematic state, so that the compensated Hall conductivity might be recovered, leading to the $B$-linear dependence of Hall resistance at low temperature. Another alternative possibility is the strain due to the interaction between ultrathin FeSe flakes and the substrates. As described in the introduction section, strains in films grown on SrTiO$_3$ and CaF$_2$ substrates have been reported to have significant effects on superconductivity and nematic state \cite{G17}, where the strain is produced by the epitaxial growth of the films and exists in all temperatures. $T_{\rm s}$ and $T_{\rm c}$ show anti-correlated relationship therein, i.e., $T_{\rm s}$ decreases accompanied with the increase of $T_{\rm c}$. While for the exfoliated flakes here, there is no strain at room temperature. The strain might emerge as temperature is cooled down because the different temperature dependence of thermal expansion coefficient between FeSe and substrates \cite{R31,M32,G33,G34}. When temperature cools from 300 to 4 K, the relative length changes along {\sl a} axis and {\sl b} axis are the same for Al$_2$O$_3$ (in the case of $d<$ 5 nm) and SiO$_2$ (in the case of $d\ge$ 5 nm)\cite{M32,G33}. According to literatures, the relative length change of Al$_2$O$_3$ is about 0.614$\times$10$^{-3}$ and SiO$_2$ is 2.44$\times$10$^{-3}$. As for FeSe, the relative length changes along orthorhombic {\sl a} axis and {\sl b} axis are different below $T_{\rm s}$ due to the tetragonal to orthorhombic structural transition. It has been reported that the relative length change along {\sl b} axis is about 3.8$\times$10$^{-3}$ with temperature cooled from 300 to 4 K and that along {\sl a} axis is negligible as comparing between 300 and 4 K \cite{R31,A36}. In this case, FeSe thin flakes could feel compressive strain along the orthorhombic {\sl a} direction while tensile strain along the orthorhombic {\sl b} direction from the substrates at low temperature, which would suppress or even wipe out the tetragonal to orthorhombic structural transition and thus the nematic transition. Such strains might have observable effects on the properties of the FeSe flakes only when the flakes are thin enough, because strain will be released in thick flakes, just as reported in the films \cite{S39}. The drop of $T_{\rm c}$ as $d <$ $\sim$ 27 nm shown in Fig. 4b might be understood that the strain starts to take effect below this thickness. Also for $d <$ $\sim$ 27 nm, the effective anisotropic strains could affect the formation of Dirac-type minor electron band with ultrahigh mobility \cite{S35} and lead to the change of $R_{xy}$ at 15 K from $B$-nonlinear to -linear dependence with decreasing the flakes thickness.  The rapid decrease of $T_{\rm s}$ and $T_{\rm c}$ might be the combining effect of decreasing size of twinned domains and appearance of the strain as $d <$ $\sim$ 27 nm. In addition, the reduction of dimensionality could play a role for the decreasing of $T_{\rm s}$ and $T_{\rm c}$ with thinning FeSe flakes.

In order to find the effect of two-dimensional superconducting fluctuations upon superconductivity, the temperature-dependent $I$-$V$ curves are measured around $T_{\rm c}^{\rm zero}$  for the ultrathin FeSe flakes(as shown Fig. 5a). As we know, the $I$-$V$ characteristic curves reveal the signature of Berezinskii-Kosterlitz-Thouless transition, which is an evidence for 2D superconductivity. According to the BKT theory, the zero-resistance state only emerges when the vortex and antivortex are bound into pairs below a so-called BKT transition temperature ($T_{\rm BKT}$) \cite{M37}. For the BKT transition, the current-induced Lorentz force causes vortex-antivortex pairs unbinding, resulting in a $V$ $\sim$ $I^{\alpha}$ behavior with $\alpha$($T_{\rm BKT}$) = 3. As shown in Fig. 5a, the measured $I$-$V$ curves of the hexalayer FeSe exhibit a $V$ $\sim$ $I^{\alpha}$ power-law dependence, and the slope corresponding to the exponent $\alpha$ changes systematically as expected for the BKT transition. We extracted the temperature dependence of the power-law exponent $\alpha$, which was deduced by fitting the $I$-$V$ curves, as shown in Fig. 5b. With decreasing temperature, the exponent $\alpha$ deviates from unity and approaches 3 at 3.52 K, which is identified as $T_{\rm BKT}$. Additionally, the temperature-dependent resistance $R({\rm T})$ follows a typical BKT-like behavior with $R({\rm T})$ = $R_{\rm 0}$exp[$-b/(T - T_{\rm BKT})^{1/2}$] in the temperature range close to $T_{\rm BKT}$, where $R_{\rm 0}$ and $b$ are material dependent parameters \cite{B38}. As shown in Fig. 5c, the extracted value of $T_{\rm BKT}$ from the measured $R({\rm T})$ curve is about 3.59 K, which is in agreement with the $I$-$V$ results.

Considering the 2D-like superconductivity in FeSe thin flake, the superconducting state should be much more robust against the in-plane magnetic field than the out-of-plane magnetic field\cite{Y41,Y42}. We measure magneto-transport in the parallel and perpendicular magnetic field orientations. Fig. 6a and 6b display the temperature-dependent resistance under different out-of-plane and in-plane magnetic fields. Such a large anisotropy suggests that the superconductivity is strongly 2D in nature. For the out-of-plane magnetic fields,superconductivity is dramatically suppressed at a small magnetic field, while superconductivity can survive even up to 9 T with the magnetic field applied along the in-plane direction. Fig. 6c shows the temperature dependence of $H_{\rm c2}$ at in-plane ($H_{\rm c2}^{\rm ab}$) and at out-of-plane($H_{\rm c2}^{\rm c}$) magnetic fields, which exhibits a good agreement with the phenomenological Ginzburg-Landau (GL) expressions for 2D superconducting films\cite{M43}. \\

$H_{\rm c2}^{\rm c}$(T) = $\frac{\Phi_0}{2\pi\xi_{GL}(0)^2}(1-\frac{T}{T_c})$\\

$H_{\rm c2}^{\rm ab}$(T) = $\frac{\sqrt{12}\Phi_0}{2\pi\xi_{GL}(0)d_{SC}}(1-\frac{T}{T_c})^{1/2}$ \\ where ${\Phi_0}$ is the magnetic flux quantum, ${\xi_{GL}(0)}$ is the in-plane 2D GL coherence length at 0 K, and $d_{SC}$ is the thickness of the superconductor. The upper critical magnetic field $H_{\rm c2}$ is estimated as the field at which the resistance equals the half value of the normal state. The temperature dependence of $H_{\rm c2}$ can be well described by the 2D standard model, indicating that perfect 2D superconducting behavior is  observed in our samples. These experimental results here support a 2D-like superconductivity in few-layer FeSe.

\section*{Conclusion}
We have successfully used the  traditionally mechanical and Al$_2$O$_3$-assisted exfoliation techniques to achieve the thin flakes FeSe with different thickness, and have systematically investigated the evolution of transport properties in FeSe thin flakes as a function of thickness. With decreasing flakes thickness, the structural transition temperature and superconducting transition temperature gradually decrease, and the $B$-nonlinear behavior for Hall resistance at low temperature is also gradually suppressed. Increase of disorder scattering from boundaries of twinned domains and anisotropic strains in the nematic state, as well as reduction of dimensionality, are considered to account for the above results. Therefore, other than the reduction of the dimensionality and the increase of disorder scattering from the boundaries of twinned domains as the FeSe flakes are thinned to 2D limitation as have reported previously \cite{L13,A27}, our results show that the anisotropic strains might also become a significant factor to tuning the electronic states in FeSe, which impacts the nematic and superconducting transition by blocking the tetragonal-orthorhombic structural transition at low temperature. Our work provides a new way to clarify the interaction between nematic and superconducting order.

\vspace*{10pt}
\noindent
\textbf{Acknowledgements}: This work was supported by the National Natural Science Foundation of China (Grant Nos. 11888101 and 11534010), the National Key Research and Development Program of the Ministry of Science and Technology of China (Grant Nos. 2017YFA0303001, 2016YFA0300201 and 2019YFA0704901), the Strategic Priority Research Program of Chinese Academy of Sciences (Grant No. XDB25000000), Anhui Initiative in Quantum Information Technologies (Grant No. AHY160000), the Science Challenge Project of China (Grant No. TZ2016004), and the Key Research Program of Frontier Sciences, CAS, China (Grant No. QYZDYSSW-SLH021).

\end{document}